\documentclass[12pt,english]{iopart} 

\usepackage[usenames,dvipsnames]{color}
\usepackage[normalem]{ulem}
\usepackage{graphicx}

\expandafter\let\csname equation*\endcsname\relax
\expandafter\let\csname endequation*\endcsname\relax

\usepackage{amsmath}
\usepackage{amssymb,amsthm}
\usepackage{amsfonts}
\usepackage{color}
\usepackage{graphics}
\usepackage{epsfig}
\usepackage{bbm}
\usepackage{pstricks}
\usepackage{subfigure}
\usepackage{bm}
\usepackage{bbold}
\usepackage{cite}

\newcommand{\be}{\begin{equation}}
\newcommand{\ee}{\end{equation}}
\newcommand{\bea}{\begin{eqnarray}}
\newcommand{\eea}{\end{eqnarray}}

\newcommand{\eins}{\mathbb{1}}
\renewcommand{\qed}{\ensuremath{\hfill \Box}}

\newcommand{\ketbra}[1]{\ensuremath{| #1 \rangle \!\langle #1 |}}

\newcommand{\ket}[1]{\ensuremath{|#1\rangle}}

\newcommand{\kommentar}[1]{}

\newcommand{\forget}[1]{}

\begin{document}

\title{Graph states and local unitary transformations beyond local 
Clifford operations}

\author{Nikoloz Tsimakuridze$^1$
and
Otfried G\"uhne$^2$}

\address{$^1$School of Mathematics and Computer Science, 
Free University of Tbilisi, \\
240 David Agmashenebeli alley, 0159 Tbilisi,
Georgia
}

\address{$^2$Naturwissenschaftlich-Technische Fakult\"at,
Universit\"at Siegen,
Walter-Flex-Stra{\ss}e~3,
57068 Siegen, Germany}


\date{\today}

\begin{abstract} 
Graph states  are quantum states
that can be described by a stabilizer formalism and play an important 
role in quantum information processing. We consider the action of local 
unitary operations on graph states and hypergraph states. We focus on 
non-Clifford operations and find for certain transformations a graphical 
description in terms of weighted hypergraphs. This leads to the 
indentification of
hypergraph states that are locally equivalent to graph states. Moreover, 
we present a systematic way to construct pairs of graph states which are
equivalent under local unitary operations, but not equivalent under local Clifford
operations. This generates counterexamples to a conjecture known as LU-LC 
conjecture. So far, the only counterexamples to this conjecture were found 
by random search. Our method reproduces the smallest known counterexample
as a special case and provides a physical interpretation.
\end{abstract}

\section{Introduction}

Multiparticle entanglement is central for many protocols in quantum information
processing. When studying its effects and properties, one often focuses on 
special families of states  with certain symmetries or a compact description. 
Two relevant families of pure multiparticle states are graph states and hypergraph 
states. Graph states are described by mathematical graphs, where the vertices 
correspond to the qubits and the edges represent particular interactions between 
them \cite{heinpra, hein}. A generalization of these states are hypergraph states 
\cite{rossinjp, guehnejpa, Qu2013_encoding, lyons, lyonsnew, gbg, chenlei,scripta}.
In a usual graph an edge connects only two vertices, 
but in a hypergraph an edge connects an arbitrary subset of vertices. It follows that 
hypergraph states are generated by 
multiparticle interactions and not only by two-particle ones. Both families of
states have found many applications: Graph states are relevant for quantum error
correction \cite{graphapp2, grassl} and measurement-based quantum computation \cite{mqc}, 
while hypergraph states occur in certain quantum algorithms \cite{scripta} and 
violate Bell inequalities in a robust way \cite{gbg}.

Entanglement is a property independent of the local basis, so it is important to
study equivalence classes under local unitary (LU) transformations. States that are
LU equivalent have the same entanglement properties and are useful for exactly the
same tasks. For graph states it has turned out that different graphs may describe 
the same state up to LU transformations. The question arises, whether such an 
equivalence can be checked easily for two given graphs. For a small number of 
qubits, it was realized that a certain discrete subset of LU transformations, 
the so-called local Clifford (LC) transformations, are the only relevant 
transformations \cite{heinpra, adan8pla, adan8pra}. Clifford transformations 
are defined by their property of leaving 
the Pauli matrices invariant. It was also shown that the action of LC transformations
on graph states corresponds to a certain transformation of the graph, called 
local complementation \cite{nestgraphicaldescription}. In this way, the 
equivalence of two graphs under LC 
transformations can easily be checked in a graphical way \cite{martenefficient}. 

The fact that LC transformations are  sufficient to characterize LU equivalence 
for up to eight qubits led to the conjecture that two graph states are LU 
equivalent, if and only if they are LC equivalent \cite{maartenlulc}. After 
formulation of this conjecture, some evidence in favor of it
has been given and it was shown to be equivalent to a property of quadratic forms
over the binary field $\mathbbm{F}^n$ \cite{maartenlulc, zengpra, nestgross}.
This LU-LC conjecture, however, is false \cite{Ji2010_LU-LC}. Zhengfeng Ji and 
coworkers found a 
counterexample for 27 qubits based on the relation of the LU-LC conjecture 
to quadratic forms. The counterexample itself was found numerically by 
random search and does not give significant insight into the question how
it may be generalised. Also, the two corresponding graphs that are 
not LC equivalent do not reveal any structure helping to understand the problem.

In this paper we clarify this situation. To do so, we study non-Clifford 
transformations of graph and hypergraph states in a systematic way. First, 
we consider the application of the transformation $U = (\sigma_x)^\alpha$ 
on a single qubit. This is an LC transformation for $\alpha = \pm 1/2$ or 
$\alpha = \pm 1$, but we derive a graphical rule for general $\alpha$ in terms 
of weighted hypergraph states. This shows that weighted hypergraph states, 
initially introduced as locally maximally entangleable (LME) states in 
Refs.~\cite{carolinapra, Carle2013}, are the natural framework to study 
this problem. Equipped with
these rules we obtain several insights: First, we give an example of a graph 
state that is LU equivalent to a hypergraph state with edges of higher cardinality.
This shows that sometimes many particle interactions are not needed for 
generating hypergraph states. Second, we present a family of pairs of graph 
states that are LU equivalent, but not LC equivalent. All these examples rely
on some bipartite structure of the underlying graph. We first discuss a 28 qubit
example, for which the main idea is easy to understand. Then, we demonstrate that the
same idea can be used to construct an example with 27 qubits. This example
is then shown to be equivalent to the example from Ref.~\cite{Ji2010_LU-LC}. 
More specifically, the example in Ref.~\cite{Ji2010_LU-LC} can be transformed 
into our example by LC operations and permutations of the qubits. Finally, we 
conclude and discuss further open
questions.

\section{Hypergraph states and weighted hypergraph states}
 
\begin{figure}
\begin{center}
\includegraphics[width=0.95\columnwidth]{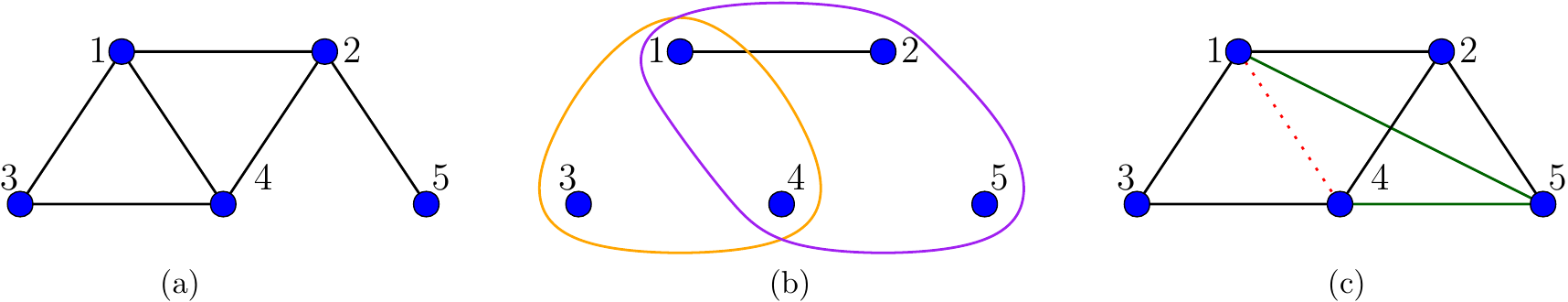}
\end{center}
\caption{(a) An example of a graph with five vertices. (b) An example of a hypergraph with five vertices. (c) Another graph of five qubits, this graph originates from the 
graph in (a) by a local complementation on the second vertex. Starting from
the second vertex one considers its neighbourhood (the vertices $1,4,$ and $5$)
and the corresponding subgraph is inverted. Existing edges ($\{1,4\}$) are 
removed, and missing ones ($\{1,5\}$ and $\{4,5\}$) are created. See text for further details.}
\label{fig:graphexamples}
\end{figure}

Let us start by introducing graph states and hypergraph states. A graph is a 
set of vertices and some edges connecting them. A hypergraph consists
also of a set of vertices and edges, but now the edges are allowed
to connect more than two vertices. Some examples of graphs and hypergraphs
are shown in Fig.~\ref{fig:graphexamples}. Of course, any graph is also a 
hypergraph, so we can write down the definitions for hypergraphs only, 
they also hold for usual graphs. More details on the general theory of 
hypergraph states can be found in Refs.~\cite{rossinjp, guehnejpa}.

Given a hypergraph $H$ with $N$ vertices we can associate to it a pure quantum state 
on $N$ qubits in the following way: Any vertex corresponds to a qubit and any qubit is prepared
in the state $\ket{+} = (\ket{0}+\ket{1})/\sqrt{2}$. Then, one applies for each edge
a multi-qubit phase gate. For an edge  $e$ containing $k$ vertices, this gate is given
by
\begin{equation}
C_e = \mathbbm{1}_e - 2 \ketbra{\underbrace{1, \dots , 1}_{k\;\; {\rm times}} }
\label{eq:phasegate}
\end{equation}
and acts on the qubits connected by $e$. The phase gate $C_e$ acts trivially on 
all other qubits. So, if we have an element of the computational basis, this vector acquires
a sign flip, if and only if it has only $1$-s on every qubit contained in $e$.
The hypergraph state is the state after application of all phase gates,\footnote{Note 
that here and in the following we sometimes use a simplified notation. First, we talk of 
the ``egdes'' of a hypergraph, although the term ``hyperedges'' would be more precise. 
Second, we write $e \in H$ to denote edges from $H$, but formally $H$ is a pair, 
consisting of a set of vertices and a set of hyperedges.
}  
\begin{equation}
\ket{H} = \big( \prod_{e \in H} C_e \big) \ket{+}^{\otimes N}.
\end{equation}
Note that all the phase gates commute, so the order of the product does not matter
here.

It is convenient that hypergraph states coincide with states which have real 
and equal weights for any member of the computational
basis \cite{rossinjp}. These states can be written as
\begin{equation}
\ket{\psi} = \frac{1}{\sqrt{2^N}} 
\sum_{x \in \{0, 1\}^N} (-1)^{f(x)} \ket{x},
\label{REWstates}
\end{equation}
where $f(x) \in \{0, 1\}$. A further very useful fact about hypergraph states is 
that they
can be described by a (non-local) stabilizer. This means that there exists an abelian
group of $2^N$ observables $S_i$ and the hypergraph state is the unique eigenstate
of all these observables, that is $S_i\ket{H} = \ket{H}$ for all $i$.  This offers an alternative definition
of hypergraph states, but this is not so important for our approach. For the case
of usual graph states, the stabilizer is local, that is, the $S_i$ are tensor products
of Pauli matrices. 

The starting point for our discussion is that different hypergraphs may describe 
the same hypergraph state up to LU transformations. Let us discuss this first for
the special case of graph states. One can ask whether for two different graphs
$G_1$ and $G_2$ the relation
\be
\ket{G_1} = U_1 \otimes U_2 \otimes \dots \otimes U_N \ket{G_2}
\ee
holds. Here, the $U_i$ are unitary transformations on the respective qubits.  
For a small number of qubits ($N\leq 8$) it has turned out that it is sufficient
to consider only special unitaries $U_i$ from the Clifford group when deciding 
LU equivalence \cite{heinpra, maartenlulc}. These unitaries leave by definition 
the set of Pauli matrices 
invariant, i.e., $U \sigma_i U^\dagger = \pm \sigma_{\pi(i)},$ where $\pi(\cdot)$
is some permutation. This is a discrete set, 
consisting of elements like $\sigma_i$ and $\sqrt{\sigma_i}$ and the Hadamard 
transformation. Moreover, the action of local Clifford (LC) operations graph 
states has a graphical interpretation in terms of a local complementation of
the graph. In this operation, a single vertex is picked and its neighbourhood is
inverted, an example is explained in Fig.~\ref{fig:graphexamples}. One can show
that this transformation corresponds to an LC transformation on the graph state 
and, conversely, any LC transformation between graph states can be written as a 
sequence of local complementations \cite{nestgraphicaldescription}. 

Of course, given this restricted set of operations with a clear interpretation, 
it is much easier to decide whether two graph states are LC equivalent.
Since LC transformations are sufficient for small systems, it is
tempting to conjecture that any LU equivalent pair of graph states is also LC 
equivalent. This is the LU-LC conjecture. This conjecture is wrong, however, and 
it is one of the main goals 
of this paper to develop a systematic procedure to generate counterexamples 
to this. Concerning LU equivalence of hypergraph states containing edges with
three or more vertices, it was shown in Refs.~\cite{guehnejpa, chenlei} that for 
$N \leq 4$ all the LU equivalent states are equivalent under a simple application
of the Pauli matrices (that is, even a smaller subset than the LC transformations),
while for $N \geq 5$ this is not the case \cite{gtg}. Still many open questions remain. 
For instance, LU transformations between graph states and hypergraph states have
not been identified so far. We will later present an example of a graph state and
a hypergraph state containing three-edges, which are LU equivalent. 

Let us finally discuss weighted hypergraph states, as they are a main tool to 
formulate our graphical rules later. Instead of the multiqubit phase gate in
Eq.~(\ref{eq:phasegate}) we consider the generalization with an arbitrary phase,
\begin{equation}
C_e^{\alpha_e} = \mathbbm{1}_e - (1-e^{i \pi \alpha_e}) 
\ketbra{\underbrace{1, \dots , 1}_{k\;\; {\rm times}} }.
\end{equation}
For $\alpha_e=1$ this is just the phase gate from above.\footnote{Note the factor 
$\pi$ in the exponent. This is used here for later convenience, but it is
not used in Refs.~\cite{carolinapra, Carle2013}.} Then, the weighted
hypergraph state is defined as 
\begin{equation}
\ket{H} = \big( \prod_{e \in H} C_e^{\alpha_e} \big) \ket{+}^{\otimes N}
\end{equation}
and it can be represented by a hypergraph where each edge carries the weight 
$\alpha_e$. We can also express the weighted hypergraph state as a (not 
necessarily real) equally weighted state 
\begin{equation}
\ket{\psi} = \frac{1}{\sqrt{2^N}} 
\sum_{x \in \{0, 1\}^N} e^{i \pi f(x)} \ket{x}.
\label{REWstates2}
\end{equation}
Contrary to Eq.~(\ref{REWstates}) the values of the function $f(x)$ are not necessarily 
in $\{0, 1\}$, but can be any real values.

These equally weighted states were originally invented as LME states due to 
their property of being maximally entangleable to auxiliary systems using 
only local operations \cite{carolinapra}. We will see that the actions of 
powers of Pauli gates to hypergraph states will in general give us such 
an LME state. Thus LME states are the fundamental objects if one wishes to
study unitary transformations of graph states or hypergraph states.

\section{The power of a single-qubit gate}
In this section we derive some facts about the power $G^\alpha$ 
of some single-qubit unitary $G$ for an arbitrary real $\alpha$.
This will be needed later for giving
the LU transformation a graphical description. 

We consider unitary operators $G$ that satisfy the condition 
$G^2 = \mathbbm{1}$. This condition is equivalent to the statement 
that $G$ is also a hermitian operator, or alternatively to the statement 
that all eigenvalues of $G$ are $\pm 1$. Such operators obey the following 
formula:
\begin{equation}
G = e^{i \pi \frac{\mathbbm{1} - G}{2}}.
\end{equation}
{From} this formula it is natural to {\it define} $G^\alpha$ for any 
real $\alpha$ as:
\begin{equation}
G^\alpha \equiv e^{i \pi \alpha \frac{\mathbbm{1} - G}{2}} 
= \frac{1}{2}
\big[(1 + e^{i \pi \alpha}) \mathbbm{1} + (1 - e^{i \pi \alpha}) G\big].
\end{equation}
Note that even for scalars non-integer powers are not uniquely defined,
thus it is natural that the same occurs with matrices. For example, 
each Pauli matrix squares to the identity matrix and thus could be 
considered to be $\mathbbm{1}^{\frac{1}{2}}$. The definition above, however, 
chooses one particular option out of all possible choices for $G^\alpha$.

This definition has some natural properties of usual exponentiation. For 
example as one would expect: $G^1 = G$ and $G^n$ defined this way coincides 
with the usual definition of integer powers of a matrix. This definition 
also coincides with the definition of $C_e^{\alpha_e}$ for weighted 
hypergraph states. Moreover, the following equation holds:
\begin{equation}
G^{\alpha_1} G^{\alpha_2} = G^{\alpha_1 + \alpha_2} \quad \mbox{ for all } \alpha_1, \alpha_2.
\end{equation}
It is important, however, to realize that in general the power of a product 
of matrices $G_1$ and $G_2$ does not always coincide with product of powers, 
even if $G_1$ and $G_2$ commute,
\begin{equation}
(G_1 G_2)^\alpha \neq G_1^\alpha G_2^\alpha.
\end{equation}
This is because our definition selects one option out of several 
for defining the power of a matrix. To give an example that is useful
later,  we consider single-qubit phase gates
\begin{equation}
C_1 = \begin{bmatrix}
1 & 0 & 0 & 0 \\
0 & 1 & 0 & 0 \\
0 & 0 & -1 & 0 \\
0 & 0 & 0 & -1
\end{bmatrix}
\quad \text{and} \quad
C_2 = \begin{bmatrix}
1 & 0 & 0 & 0 \\
0 & -1 & 0 & 0 \\
0 & 0 & 1 & 0 \\
0 & 0 & 0 & -1
\end{bmatrix},
\end{equation}
acting on a two-qubit system. Then the product of their powers and the power 
of their products can be expressed as follows:
\begin{equation}
C_1^{\alpha} \cdot C_2^{\alpha} = \begin{bmatrix}
1 & 0 & 0 & 0 \\
0 & e^{i \pi \alpha} & 0 & 0 \\
0 & 0 & e^{i \pi \alpha} & 0 \\
0 & 0 & 0 & e^{i \pi 2 \alpha}
\end{bmatrix}
\quad
(C_1 \cdot C_2)^{\alpha} = \begin{bmatrix}
1 & 0 & 0 & 0 \\
0 & e^{i \pi \alpha} & 0 & 0 \\
0 & 0 & e^{i \pi \alpha} & 0 \\
0 & 0 & 0 & 1
\end{bmatrix}.
\end{equation}
Thus we need the two-qubit phase gate  $C_{\{1, 2\}}^{- 2 \alpha}$ as a
correction term to obtain  $(C_1 C_2)^\alpha$ from $C_1^\alpha \cdot C_2^\alpha$, 
where
\begin{equation}
C_{\{1, 2\}} = \begin{bmatrix}
1 & 0 & 0 & 0 \\
0 & 1 & 0 & 0 \\
0 & 0 & 1 & 0 \\
0 & 0 & 0 & -1
\end{bmatrix}
\quad \text{and thus} \quad
C_{\{1, 2\}}^{- 2 \alpha} = \begin{bmatrix}
1 & 0 & 0 & 0 \\
0 & 1 & 0 & 0 \\
0 & 0 & 1 & 0 \\
0 & 0 & 0 & e^{- i \pi 2 \alpha}
\end{bmatrix}.
\end{equation}
In this example, we have seen that powers of single-qubit operators
naturally lead to weighted multi-qubit unitaries. In the following 
section, we will see a more general formula for powers of such 
product operators.
 
\section{A graphical rule for the action of $X^\alpha$ on a 
hypergraph state}
In this section, we formulate graphical rules for the action of 
some local unitaries on hypergraph states. We consider powers of
the Pauli matrices $\sigma_x, \sigma_y$ and $\sigma_z.$ {From} now
on, we will abbreviate the Pauli matrices simply by $X, Y$, and $Z$.

First, if the Pauli $Z$ gate acts on qubit $i$ with some power 
$\alpha$, the action of $Z_i^\alpha = C_i^{\alpha}$ to the hypergraph state is 
easy to describe in terms of a weighted hypergraph. Such gates 
just add weight $\alpha$ to the edge $e =\{ i \}$ that contains 
the single qubit $i$. On the other hand, the Pauli gate $Y^\alpha$, 
in general, does not transform hypergraph states into hypergraph 
states, it leads out of the space of weighted hypergraph states.

So let us focus on the Pauli $X$ gate. With the power $\alpha=1$ it 
always transforms hypergraph states into hypergraph states and its 
effects are well known \cite{guehnejpa, Qu2013_encoding}. Specifically, 
applied to the $i$-th qubit of the hypergraph state 
$\ket{H} = ( \prod_{e \in H} C_e ) \ket{+}^{\otimes N}$ 
it produces 
the state $C_{\Delta_i H} \ket{H}$ where $C_{\Delta_i H}$ is a 
diagonal 
unitary operator defined as follows: $C_{\Delta_i H}$ causes the appearance (or disappearance 
if they 
are already present) of all edges in the hypergraph $\Delta_i H$. 
The hypergraph $\Delta_i H$ is formed by taking all edges in 
$H$ that contain the vertex $i$ and removing the vertex $i$ 
from each of them,
\begin{align}
\Delta_i H &= \{ e - \{ i \} \;|\; i \in e, e \in H \}
\nonumber
\\
&= \{e' \subseteq \{1, \dots N \} \; |\;  i \notin e', e' \cup \{ i \} \in H \},
\\
C_{\Delta_i H} &= \prod_{e \in \Delta_i H} C_e.
\end{align}
The proof of this rule follows from the commutation relations between
the $X_i$ and the phase gates $C_e$, see Ref.~\cite{guehnejpa} for many
examples of this rule. 

Let us now consider the action of the $X_i^\alpha$ gate on a hypergraph 
state $\ket{H}$. 
Since $X_i^\alpha$ can be decomposed into a weighted sum of $\eins$ 
and $X_i$ gates, its action on $\ket{H}$ state is easy to characterize:
\begin{align}
X_i^\alpha \ket{H} &= X_i^\alpha C_H \ket{+}^{\otimes N}
\nonumber \\
&= \frac{1}{2} 
\big[
(1 + e^{i \pi \alpha}) \eins + (1 - e^{i \pi \alpha}) X_i
\big] 
C_H \ket{+}^{\otimes N} 
\nonumber
\\
&= \frac{1}{2} C_H \big[(1 + e^{i \pi \alpha}) \eins + (1 - e^{i \pi \alpha}) C_{\Delta_i H} 
X_i\big] \ket{+}^{\otimes N} 
\nonumber
\\
&= \frac{1}{2} C_H 
\big[(1 + e^{i \pi \alpha}) \eins + (1 - e^{i \pi \alpha}) C_{\Delta_i H}\big] \ket{+}^{\otimes N} 
\nonumber
\\
&= C_{\Delta_i H}^\alpha C_H \ket{+}^{\otimes N} = C_{\Delta_i H}^\alpha \ket{H}.
\end{align}
Thus we see that the result of $X_i^\alpha$ acting on $\ket{H}$ is the 
same as the result of $C_{\Delta_i H}^\alpha$ acting on $\ket{H}$,
\begin{equation}
(X_i)^\alpha \ket{H} = (C_{\Delta_i H})^\alpha \ket{H}.
\end{equation}
To understand the graphical interpretation of this action, we need to 
look at $C_{\Delta_i H}^\alpha$ and decompose it into actions of gates 
that change the weight of single hyperedges. In general,
$C_{\Delta_i H}$ is a product of several  hyperedge gates,
$C_{\Delta_i H} = C_{e_1} C_{e_2} \ldots C_{e_k}$, and for taking the
power we have to recall that $(G_1 G_2)^\alpha \neq G_1^\alpha G_2^\alpha$. 
Thus we need to study the decomposition of powers of products of 
individual edge producing gates.

To calculate $C_{\Delta_i H}^\alpha$, we need to look at the 
eigenspace of $C_{\Delta_i H}$ with eigenvalue $-1$ and change 
its eigenvalue to $e^{i \pi \alpha}$. Since $C_{\Delta_i H}$ 
is diagonal in the computational basis, this eigenspace is 
spanned by vectors in the computational basis. In fact, it is 
spanned by all vectors in the computational basis that are 
contained in the symmetric difference (denoted by $\bigoplus$) 
of the $-1$ eigenspace basis-vectors of all the edges in $\Delta_i H$.
We can compute the indicator function ${\rm \bf I}_{\bigoplus_{i=1}^k e_i}$ of 
the symmetric difference of $k$ sets using a formula similar 
to the inclusion-exclusion principle,
\begin{equation}
{\rm \bf I}_{\bigoplus_{i = 1}^k e_i} = 
\sum_{S \subseteq \{ 1, 2 \ldots k \}, S \neq \emptyset }
(-2)^{|S| - 1} {\rm \bf I}_{\bigcup\limits_{j \in S} e_j}.
\end{equation}
Using this formula, we can obtain an expression for powers 
of products of $C_e$ gates. The simplest case was already discussed
in the previous section:
\begin{align}
(C_{e_1} C_{e_2})^\alpha 
&= C_{e_1}^\alpha C_{e_2}^\alpha C_{e_1 \cup e_2}^{-2 \alpha},
\\
(C_{e_1} C_{e_2} C_{e_3})^\alpha 
&= C_{e_1}^\alpha C_{e_2}^\alpha C_{e_3}^\alpha 
C_{e_1 \cup e_2}^{-2 \alpha} C_{e_1 \cup e_3}^{-2 \alpha} C_{e_2 \cup e_3}^{-2 \alpha}
C_{e_1 \cup e_2 \cup e_3}^{4 \alpha} ,
\end{align}
and generally 
\begin{align}
\big(\prod\limits_{i=1}^{k} C_{e_i}\big)^\alpha 
& = 
\prod\limits_{S \subseteq \{1, 2 \ldots k \}, S \neq \emptyset} 
(C_{\bigcup\limits_{j \in S} e_j})^{(-2)^{|S| - 1} \alpha}.
\end{align}
Thus the effect of the $X^\alpha$ gate on qubit $i$ in a hypergraph 
state $\ket{H}$ is the following:
\begin{itemize}
\item For each edge $e'$ in 
$\mathcal{A}_i = \{e' \subset \{1, \dots, N \}| i \notin e', e' \cup \{i\} \in H \}$ 
add weight $\alpha$ to the edge $e'$ in the hypergraph.
\item For each pair of edges $\{ e_1, e_2 \in \mathcal{A}_i | e_1 \neq e_2 \}$ 
subtract weight $2 \alpha$ from edge $e_1 \cup e_2$
\item For each triplet of distinct edges 
$\{ e_1, e_2, e_3 \in \mathcal{A}_i \}$ add weight 
$4 \alpha$ to the edge $e_1 \cup e_2 \cup e_3$
\item In general: For each $k$-tuple of distinct edges in 
$e_1 \ldots e_k \in \mathcal{A}_i$ add weight $(-2)^{k-1} \alpha$ 
to the edge $\cup_{j = 1}^k e_j.$
\end{itemize}

\begin{figure}[t]
\begin{center}
\includegraphics[width=0.9\columnwidth]{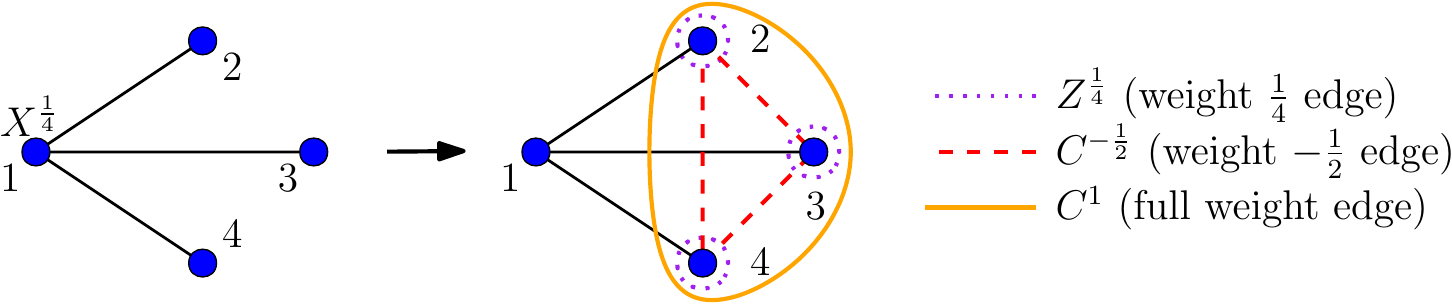}
\end{center}
\caption{An example for the graphical action of the 
$X^\alpha$ gate. The action of $X^\frac{1}{4}$ gate 
on the first qubit of the four-qubit star graph
produces three single-qubit edges $\{2\}, \{3\}$ 
and $\{4\}$ with weights $\frac{1}{4}$, three two-qubit 
edges $\{2, 3\}, \{2, 4\}$ and $\{3, 4\}$ with weight 
$-\frac{1}{2}$ and a single full-weight three-qubit 
hyperedge $\{2, 3, 4\}$.
}
\label{fig:gate-action-example}
\end{figure}

An example of these rules is shown in Fig.~\ref{fig:gate-action-example}.
We can easily see that if $\alpha = 1$ all the terms with unions 
of two or more edges add weight that is a multiple of two. Since 
$C_e^2=\eins$ we only need to consider the weight modulo two, so 
these terms act trivially on the hypergraph. The remaining terms 
correctly describe the action of Pauli $X$ gate on the hypergraph.

When $\alpha = \frac{1}{2}$, all the terms with triple or larger 
unions add weight that is a multiple of two and thus act trivially. 
Therefore only terms with single edges and pairs of edges remain. 
If we look at the effect graphically we can see that this corresponds 
(up to some single-vertex actions of weight $1/2$) to the local 
complementation operation of the graph state around vertex $i$ (see Fig.~\ref{fig:graphexamples}). This is again to be expected, since 
$X^\frac{1}{2}$ is (up to some single qubit terms $Z^{\frac{1}{2}}$)
the gate that performs local complementation for graph states.

For other values of $\alpha$ we can obtain interesting examples 
of transformations of graph and hypergraph states. Some of these can be obtained using 
Clifford gates only, but some others require the use of non-Clifford gates. Such 
examples, along with proofs that they cannot be obtained 
using Clifford transformations, are described in following sections.

\section{Graph states and hypergraph states can be LU equivalent}
In order to see how our graphical rule is useful, we now  describe
an example of a hypergraph state that is LU equivalent to a normal
graph state. The trick that we use is later relevant for constructing
counterexamples to the LU-LC conjecture. 

\begin{figure}[t]
\begin{center}
\includegraphics[width=0.9\columnwidth]{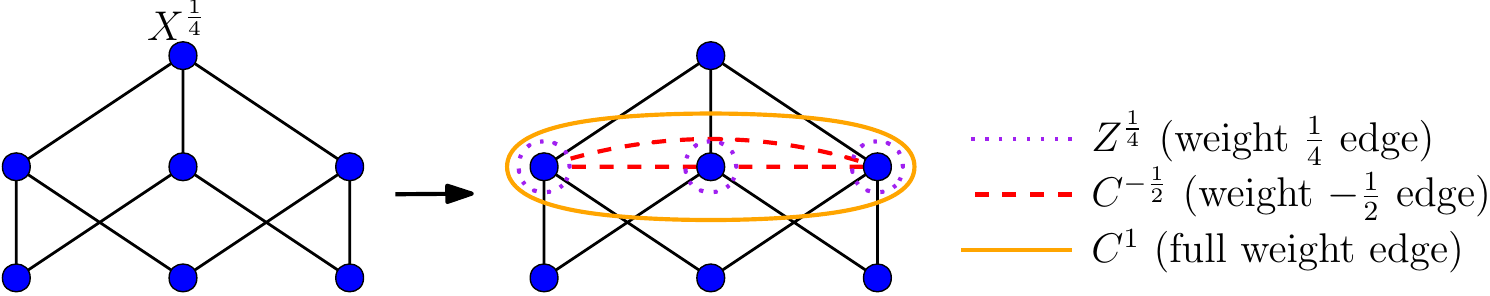}
\end{center}
\caption{The first stage of transformation of a pure graph 
state into a hypergraph state with a three-edge. See text 
for further details.
}
\label{fig:producing-hyperedge-1}
\end{figure}

We have seen already in Fig.~\ref{fig:gate-action-example}
how a $X^\frac{1}{4}$ gate can create a three-edge acting on a graph state,
but we also created some other edges with fractional weights. To obtain 
a pure hypergraph state with no fractional weight edges, we will need 
to cancel the fractional edges by other fractional power gates. 
Consider an example where we have a four-qubit star graph as before, 
but now each pair of non-central qubits is also connected to a different 
vertex. As shown on Fig.~\ref{fig:producing-hyperedge-1}, if we apply 
$X^\frac{1}{4}$ on the central qubit of the star graph, we obtain a 
three-hyperedge with several fractional two-edges and one-edges.

\begin{figure}[t]
\begin{center}
\includegraphics[width=0.9\columnwidth]{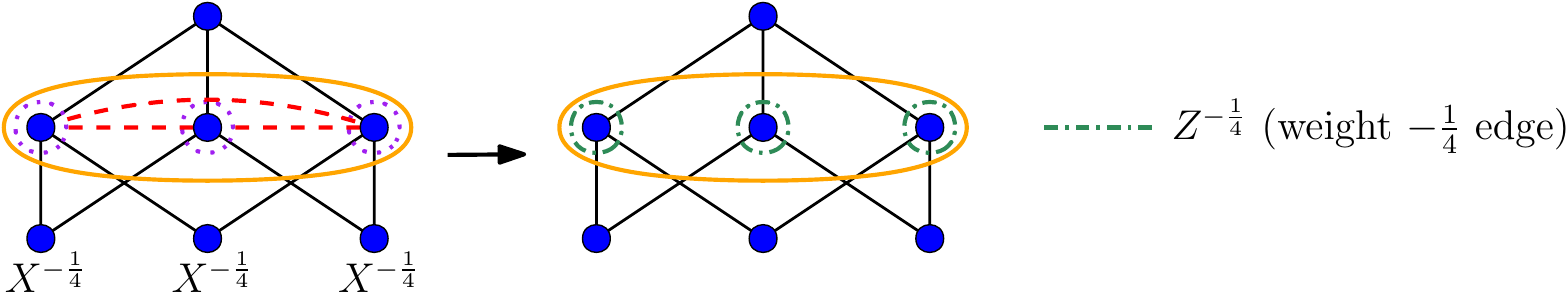}
\end{center}
\caption{In the second stage, one has to cancel the partial two-edges 
that appeared after the first stage. This is done by applying 
$X^{-\frac{1}{4}}$ gates to the qubits on the bottom.
}
\label{fig:producing-hyperedge-2}
\end{figure}

However, as demonstrated in Fig.~\ref{fig:producing-hyperedge-2} 
we can cancel the partial two-edges by applying $X^{-\frac{1}{4}}$ 
on the three qubits that connected to pairs of affected vertices. 
These gates will cancel the two-edges, but since two of them 
add weight $-\frac{1}{4}$ to every single-qubit edge, each 
single-qubit edge weight will reverse from $\frac{1}{4}$ 
to $-\frac{1}{4}.$

\begin{figure}
\begin{center}
\includegraphics[width=0.6\columnwidth]{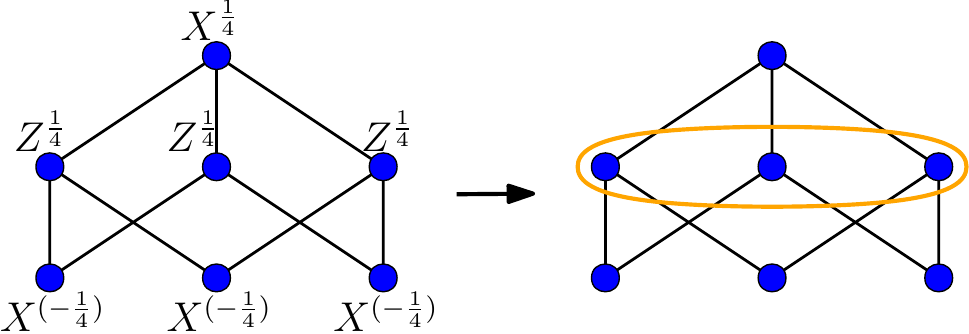}
\end{center}
\caption{Complete transformation of a pure graph state into a hypergraph 
state via local unitary transformations.}
\label{fig:graph-into-hypergraph}
\end{figure}

This leaves us nearly with a hypergraph state, we only need to cancel 
the fractional single qubit edges. For this we can just use local 
$Z^\frac{1}{4}$ gates on those qubits and obtain a hypergraph state. 
Fig.~\ref{fig:graph-into-hypergraph} combines all three steps and 
shows how a graph state can be transformed into a hypergraph state 
with a three-edge. Since LC actions cannot transform
a graph state into a hypergraph states, this gives us an example of two 
hypergraph states that are LU equivalent, but not LC equivalent.

\section{Generating counterexamples to the LU-LC conjecture}

In this section we will show how to systematically generate pairs
of graph states that are LU equivalent, but not LC equivalent. These
are then counterexamples to the LU-LC conjecture. We will show how 
to construct an example with $28$ qubits, but our example generalises 
to more qubits. In the following section, we will also construct a 
counterexample with 27 qubits, but the construction with 28 qubits is
simpler, so we explain it first. 

\begin{figure}
\begin{center}
\includegraphics{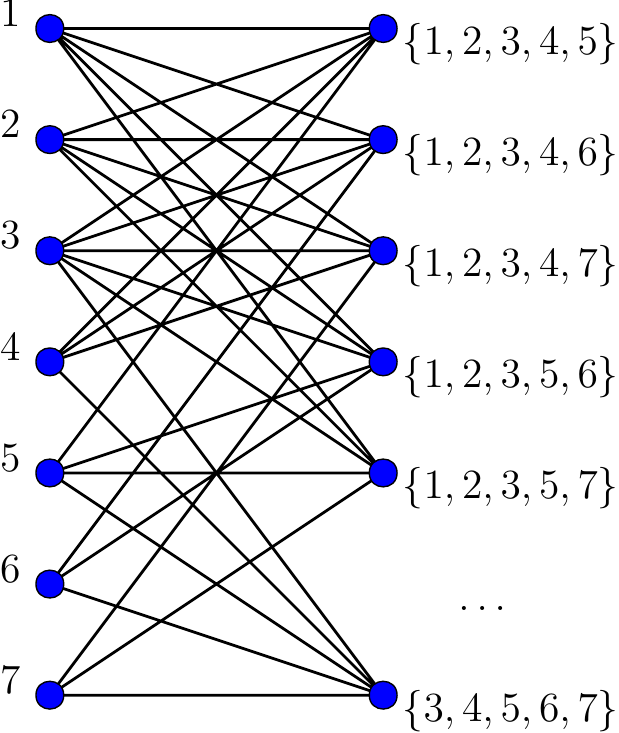}
\end{center}
\caption{A bipartite graph construction with seven vertices on the left side and 
$21=\binom{7}{5}$ vertices on the right side. Each of the vertices on the right 
side is connected to a five-vertex subset of the vertices on the left side. This 
construction is used to find counterexamples to the LU-LC conjecture, see 
text for further details.}
\label{fig:bipartite-construction}
\end{figure}

Consider the graph $G_1$ from Fig.~\ref{fig:bipartite-construction}.
This is a bipartite graph with seven vertices on one side (the ``left'' side) 
and $\binom{7}{5} = 21$ vertices on the other side (the ``right'' side). 
Each vertex on the right side corresponds to a set of five vertices 
on the left side and is connected with an edge to exactly those five 
vertices. For a possible generalization, an analogous construction with 
$n = 8 k - 1$ vertices on the left side and $\binom{n}{5}$ vertices 
on the right side works as well. We can also obtain a similar construction
if we take $n = 8 k - 1$ vertices on the left side and $\binom{n}{4}$ vertices 
on the right, each of which is connected to a unique set of four vertices 
on the left. But for the sake of simplicity we consider the graph from 
Fig.~\ref{fig:bipartite-construction} in the following.

Let us see what happens if we perform $X^\frac{1}{4}$ on every vertex 
on the right side:

\begin{itemize}

\item Each single-qubit edge on the left side will appear with weight 
$\frac{1}{4}$, multiplied with the number of neighbours the qubit has (which is 
$\binom{6}{4} = 15$). Thus we will have all single-qubit edges with 
weights of $\frac{15}{4}$, but we can easily cancel these by applying 
the $Z^{-\frac{15}{4}}$ gates to every qubit on the left side.

\item Each two-edge between vertices on the left side
will appear with weight $-\frac{1}{2}$, multiplied with 
the number of qubits on the right that are connected to both of the 
ends of such an edge. The number of such qubits on the right is 
$\binom{7 - 2}{5 - 2} = \binom{5}{3} = 10$, thus every edge will 
appear with weight $-5$, which is equivalent to $1$ modulo $2$. 
Thus we will make every possible two-edge on the left side appear.

\item Each three-edge will appear with weight $1$, multiplied with
the number 
of qubits on the right that are connected to all three qubits in the 
edge. There are $\binom{7 - 3}{5 - 3} = \binom{4}{2} = 6$ such qubits. 
Thus the three-edge will appear with weight $6$, which is equivalent 
to not appearing at all, since the weights are counted modulo $2$.

\item All four-edges and five-edges appear with weights that are 
multiples of $2$, which is also equivalent to not appearing at all.

\end{itemize}

Thus the graph state of the bipartite graph $G_1$ described above 
is LU equivalent to the graph state of graph $G_2$, which has the 
same edges between the two parts as $G_1$ and also has no edges 
between vertices on the right side but, unlike $G_1$, all the 
vertices of $G_2$ on the left side are connected to each other.

Now we need to show that $G_2$ cannot be obtained from $G_1$ by 
local Clifford operations only. It has been shown \cite{nestgraphicaldescription} 
that any Clifford transformation can be decomposed into a series of local 
complementation operations on the graph state. Using this fact, we break 
our proof into two parts.

\begin{enumerate}
\item First we show that $G_2$ cannot be obtained from $G_1$ using 
only local complementation operations on vertices in the right hand 
side of the graph.

\item Then we show a general Lemma, stating that if we can obtain a 
graph state $G_2$ from a bipartite graph state $G_1$, and $G_2$ only 
differs from $G_1$ by edges between vertices on the left side, then 
it must be the case that local complementations of the right side 
suffice to produce the transformation. Together with the first part 
this will complete the proof.
\end{enumerate}

\subsubsection*{First part of the proof. ---}

The first part of the proof is quite simple. We only need to observe 
that each local complementation operation on vertices in the right-hand 
side affects $\binom{5}{2} = 10$ edges. Since that is an even number, 
the parity of the number of edges on the left always remains even (as 
it started from the even number zero). Thus, it must be even at the 
end of the transformation. However, $\binom{7}{2} = 21$ is an odd 
number. This shows that we can never obtain all edges on the left-hand 
side using only local complementations on the right side.

\subsubsection*{Second part of the proof. ---}

In this part we to show that if the graph state $\ket{G_2}$ can be 
obtained from $\ket{G_1}$ using Clifford operations, then there must 
be a set of vertices on the right side, such that $G_1$ is transformed 
into $G_2$ by performing local complementation operations on exactly 
those vertices. We will show this by proving a more general fact
that works for almost any bipartite $G_1$, that is, any graph $G_1$ 
that can be divided into two sides with edges running only from one 
side to another and never within one side. We will also use the 
fact that there is a path from every vertex to every other vertex 
in $G_1$, which means that $G_1$ is connected. The mathematical 
result is the following:

\noindent
{\bf Lemma.}
{\it
Let $G_1$ be a connected bipartite graph that has $k_1$ vertices on the left
side of the bipartition and $k_2$ on the right side, with $k_1 \neq k_2$. Let 
$G_2$ be a graph with the same vertices and the  same edges between the two 
sides, but some extra edges added on the left side of the graph. Then, if the
graph state $\ket{G_1}$ can be transformed into graph state $\ket{G_2}$ using 
local Clifford gates, the graph $G_1$ can be transformed into $G_2$  
using local complementation applied to vertices on the right side only.}

{\it Proof.}
It has been shown in Ref.~\cite{nestgraphicaldescription} that whether 
two graph states are equivalent under LC operations can be determined 
using a set of equations over the binary field. Specifically, let us 
consider the stabiliser matrices $S_1$ and $S_2$ of the corresponding 
graphs $G_1$ and $G_2$ as $2n \times n$ matrices, where $n$ is the total 
number of vertices in each graph. These matrices have the adjacency 
matrix of the corresponding graph in the top part and the identity matrix 
in the bottom part, i.e.,
\begin{equation}
S_1 = \begin{bmatrix}
\theta_1 \\
\mathbbm{1}
\end{bmatrix},
\quad
S_2 = \begin{bmatrix}
\theta_2 \\
\mathbbm{1}
\end{bmatrix},
\end{equation}
with $\theta_1$ and $\theta_2$ being the adjacency matrices of the 
corresponding graphs. Then the graph states $\ket{G_1}$ and $\ket{G_2}$ 
are LC equivalent if and only if there exists a binary matrix $Q$ of 
size $2n \times 2n$ with a block structure
\begin{equation}
Q = \begin{bmatrix}
A & B \\
C & D
\end{bmatrix},
\end{equation}
where $A$, $B$, $C$ and $D$ are diagonal and satisfy the following 
equations:
\begin{align}
A_{ii} D_{ii} + B_{ii} C_{ii} = 1 \quad \mbox{ for all } i
\label{nondegenerate-transform-condition}
\end{align}
and
\begin{align}
S_1^T \cdot Q^T \cdot \begin{bmatrix}
0 & \eins \\
\eins & 0
\end{bmatrix} \cdot S_2 = 0.
\label{clifford-equivalence-condition}
\end{align}

Substituting $S_1$ and $S_2$ into Eq.~(\ref{clifford-equivalence-condition}) 
we obtain the following criterion:
\begin{equation}
\theta_1 C \theta_2 + \theta_1 A + D \theta_2 + B = 0.
\label{graph-clifford-equivalence}
\end{equation}
In our case the first graph $G_1$ is bipartite and the second one is obtained 
from the first one by adding some edges on one side of the graph. Thus the 
matrices $\theta_1$ and $\theta_2$ have the following special forms:
\begin{equation}
\theta_1 = \begin{bmatrix}
0 & \zeta \\
\zeta^T & 0
\end{bmatrix},
\quad
\theta_2 = \begin{bmatrix}
\eta & \zeta \\
\zeta^T & 0
\end{bmatrix},
\end{equation}
where $\eta$ is the adjacency matrix of the subgraph on the left 
side generated by the transformation, and $\zeta$ is an $k_1 \times k_2$ 
matrix that shows which vertices on the left side are adjacent to 
which ones on the right side.
We also use the fact that, since all of $A$, $B$, $C$ and $D$ are diagonal, 
they can be written as:
\begin{equation}
A = \begin{bmatrix}
A_u & 0 \\
0 & A_l
\end{bmatrix},
\quad
B = \begin{bmatrix}
B_u & 0 \\
0 & B_l
\end{bmatrix},
\quad
C = \begin{bmatrix}
C_u & 0 \\
0 & C_l
\end{bmatrix},
\quad
D = \begin{bmatrix}
D_u & 0 \\
0 & D_l
\end{bmatrix},
\end{equation}
with all the nonzero upper parts sized $k_1 \times k_1$ and all the nonzero 
lower parts sized $k_2 \times k_2$.

Then, Eq.~(\ref{graph-clifford-equivalence}) gives us the following 
condition:
\begin{equation}
\begin{bmatrix}
\zeta C_l \zeta^T & 0 \\
\zeta^T C_u \eta & \zeta^T C_u \zeta
\end{bmatrix}
+
\begin{bmatrix}
0 & \zeta A_l \\
\zeta^T A_u & 0
\end{bmatrix}
+
\begin{bmatrix}
D_u \eta & D_u \zeta \\
D_l \zeta^T & 0
\end{bmatrix}
+
\begin{bmatrix}
B_u & 0 \\
0 & B_l
\end{bmatrix}
=
\begin{bmatrix}
0 & 0 \\
0 & 0
\end{bmatrix}.
\end{equation}
Thus we have four matrix equations over the binary field:
\begin{align}
\zeta C_l \zeta^T + D_u \eta + B_u = 0,
\label{top-left-condition} 
\\
\zeta A_l + D_u \zeta = 0,
\label{top-right-condition} 
\\
\zeta^T C_u \eta + \zeta^T A_u + D_l \zeta^T = 0,
\label{bottom-left-condition} 
\\
\zeta^T C_u \zeta + B_l = 0.
\label{bottom-right-condition}
\end{align}

Let us analyse Eq.~(\ref{top-right-condition}) first. Since both $A_l$ 
and $D_u$ are diagonal, the terms $\zeta A_l$ and $D_u \zeta$ correspond 
to some selections of columns and rows of $\zeta$ respectively. These 
two terms need to sum to zero, which is the same as being equal over 
the binary field. But for every $i$, such that $(A_l)_{ii} = 1$ we 
have a full column $i$ of $\zeta$ in $\zeta A_l$, and thus for every 
position $j$ where $\zeta_{ij} = 1$ we must also have the $j$-th 
row in $D_u \zeta$. Therefore, we need to have $(D_u)_{jj} = 1$. 
In graph theoretic terms, whenever we have $(A_l)_{ii} = 1$ we 
need to have $(D_u)_{jj} = 1$ for every vertex $j$ that is 
connected to $i$. Similarly, whenever we have $(D_u)_{jj} = 1$, 
we must have $(A_l)_{ii} = 1$ for every vertex $i$ connected to 
$j$. But since the whole graph is connected, the matrices $A_l$ and $D_u$ 
must simultaneously both be zero matrices or both be identity 
matrices of their respective dimensions.
Hence, we have two cases to consider:

{\it Case 1:}
If $A_l = 0_{k_2 \times k_2}$ and $D_u = 0_{k_1 \times k_1}$ the term 
$A_{ii} D_{ii}$ is zero for all $i$ both in the upper and in the lower 
parts of the matrix. Thus $B_{ii} C_{ii}$ has to be one everywhere, 
meaning that $B_u = C_u = \mathbbm{1}_{k_1 \times k_1}$ and 
$B_l = C_l = \mathbbm{1}_{k_2 \times k_2}$. Then Eq.~(\ref{top-left-condition})
gives us $\zeta \zeta^T = \mathbbm{1}_{k_1 \times k_1}$, and Eq.~(\ref{bottom-right-condition}) gives us $\zeta^T \zeta = \mathbbm{1}_{k_2 \times k_2}$.
But this cannot be: $\zeta$ and $\zeta^T$ are rectangular matrices and 
cannot have rank more than the $\min(k_1, k_2)$. This means that 
$\zeta^T \zeta$ and $\zeta \zeta^T$ must also have rank no higher 
than $\min(k_1, k_2)$ and since $k_1 \neq k_2$ this contradicts 
at least one of the equations $\zeta \zeta^T = \mathbbm{1}_{k_1 \times k_1}$ 
or $\zeta^T \zeta = \mathbbm{1}_{k_2 \times k_2}$. 
Thus this Case~1 cannot happen.

{\it Case 2:}
If $A_l = \mathbbm{1}_{k_2 \times k_2}$ and $D_u = \mathbbm{1}_{k_1 \times k_1}$ 
the Eq.~(\ref{top-left-condition}) simplifies to the condition
$\zeta C_l \zeta^T = \eta + B_u$. Let us look what happens to the adjacency 
matrix if we apply local complementation on vertex $j$ on the right side. 
This changes the matrix $\eta$. As we know, we affect edges that connect 
pairs of vertices neighbouring vertex $j$. Thus the part $\eta$ of the 
adjacency matrix will change by the matrix $\zeta O_j \zeta^T$ (up to 
a diagonal correction), where $O_j$ is a matrix with zeros everywhere 
except on the $j$-th entry of the diagonal, which contains a $1$.

Let us take the set of vertices $j$ on the right side, such that 
$(C_l)_{jj} = 1$. If we start from graph $G_1$ and apply local 
complementation to exactly those vertices, then we produce the 
following adjacency matrix on the left:
\begin{equation}
\zeta O_{j_1} \zeta^T + \zeta O_{j_2} \zeta^T + \ldots + \zeta O_{j_m} \zeta^T 
= \zeta C_l \zeta^T.
\end{equation}
Thus we obtain the adjacency matrix $\eta$ (up to a diagonal correction, 
parametrized by $B_u$) using local complementations on the vertices $j$ 
with $(C_l)_{jj} = 1$ the right side. Since local complementations produce 
graph states from graph states the diagonal entries of $\eta$ and our 
resulting adjacency matrix will also match. Thus as we wanted to show,
if there is any LC transformation that produces the graph state with
the desired form, then local complementations on vertices in the right 
side suffice for such a transformation.
\qed

This theorem completes the proof that the examples of LU equivalent
graph states from above cannot be transformed into each other by 
means of LC transformations. Note that we did not need the lower 
right side of the adjacency matrix $\theta_1$ to be zero. Thus 
our proof could be generalised to the case when the initial graph 
is not bipartite, but has some edges connecting right side qubits 
to each other. However we still need to require that the final 
graph has no edges connecting right side qubits to each other. 
In this case we would see that any local Clifford transformation 
could be achieved by first cancelling the edges on the right side 
using local complementations on the left and then creating the 
edges on the left side by using local complementations on the 
right. However, our construction does not need such a generalised 
version of the theorem.

\section{A counterexample to the LU-LC conjecture with 27 qubits}
So far, we have presented a family of counterexamples to the LU-LC 
conjecture, the smallest one having 28 qubits. In the literature, 
however, the smallest explicit counterexample has 27 qubits and has
been found by numerical search \cite{Ji2010_LU-LC}. So, one may 
ask whether our construction and the known example are related. 

\begin{figure}
\begin{center}
\includegraphics{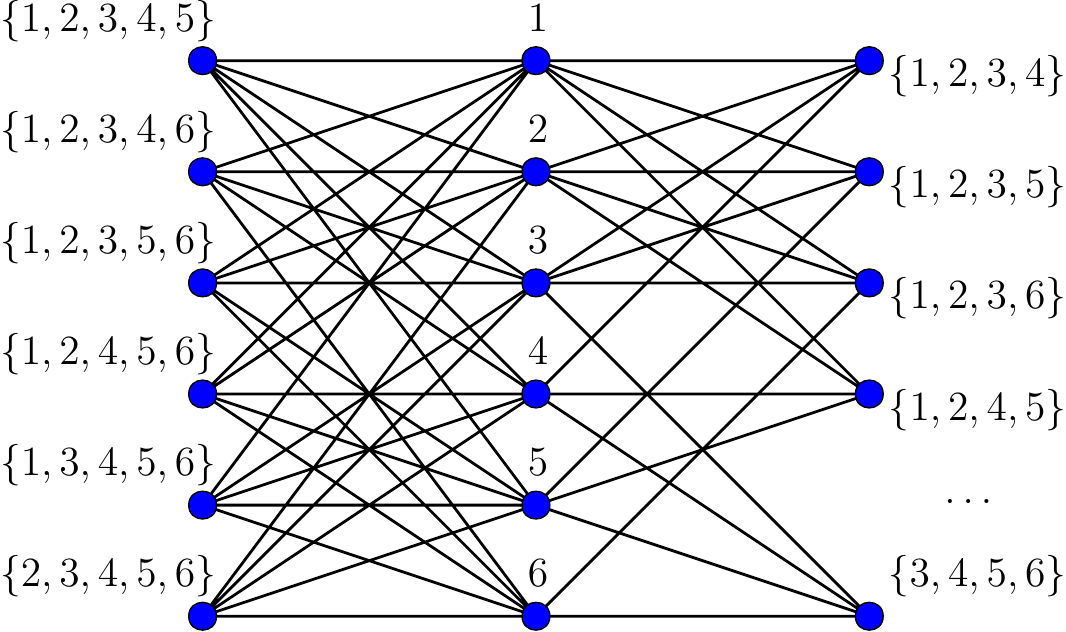}
\end{center}
\caption{The counterexample to the LU-LC conjecture for 27 qubits. 
The graph has six central vertices, 
$6=\binom{6}{5}$ vertices on the right-hand side and 
$15 = \binom{6}{4}$ vertices on the left side. Applying
$X^{\frac{1}{4}}$ on all qubits on the right side and on the 
left side induces all possible connections between the 
central vertices, i.e. the central subgraph becomes fully 
connected. This cannot be achieved by local complementation. 
}
\label{fig:tripartite-construction}
\end{figure}

\begin{figure}
\begin{center}
\includegraphics[width=0.81\columnwidth]{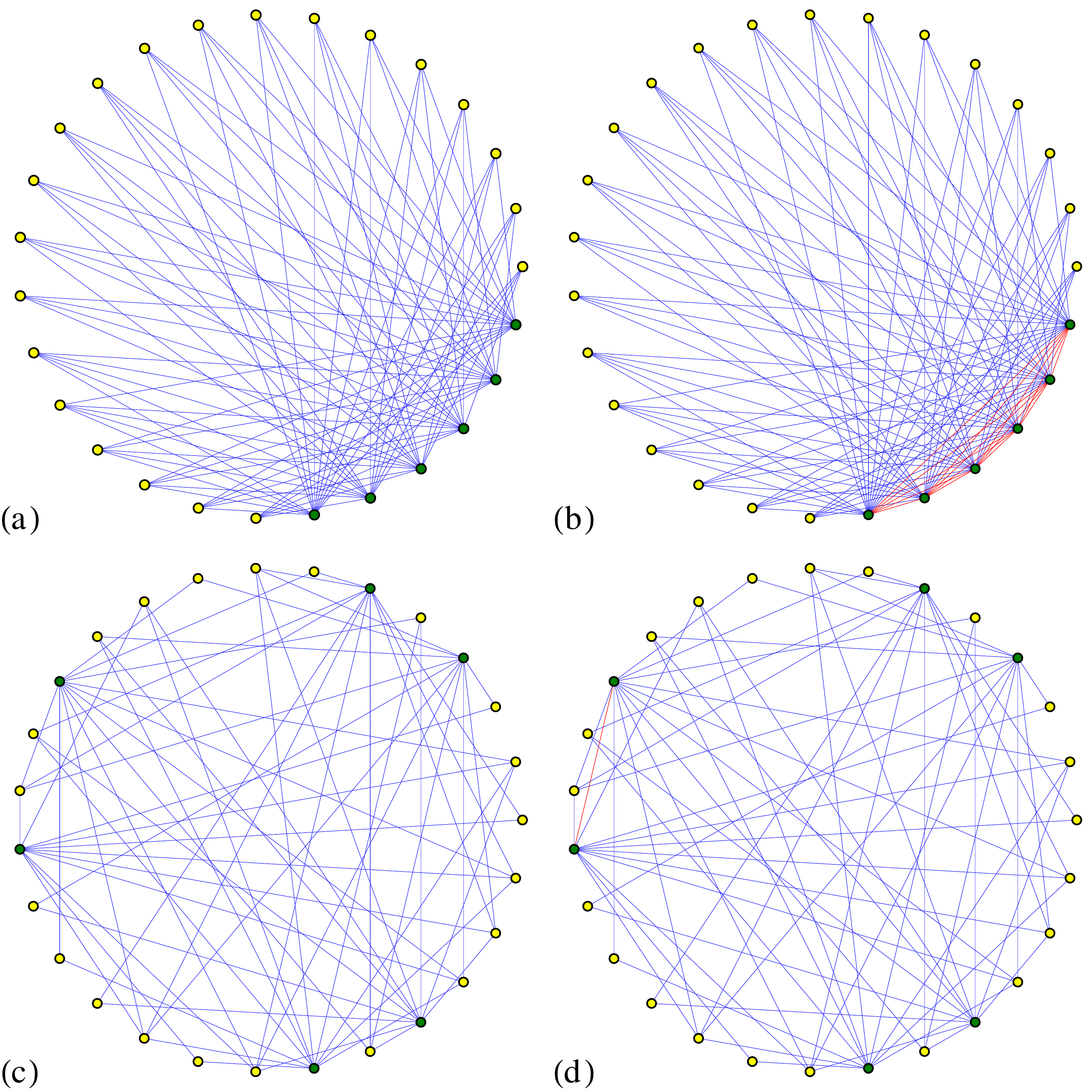}
\end{center}
\caption{The relation between the different counterexamples to the LU-LC
conjecture for 27 qubits. 
{(a)} The graph from Fig.~\ref{fig:tripartite-construction}
\label{fig:counterexamples27} which forms the starting point of our 
considerations. This is a bipartite graph, the bipartition is highlighted
by the yellow and green vertices.
{(b)} After applying LU transformations to the
graph in (a), the central qubits become connected (shown in red). 
{(c)} The graph from Ref.~\cite{Ji2010_LU-LC}, which was used 
there as a starting point. 
This is also a bipartite graph, the bipartition is highlighted 
by the yellow and dark green vertices. 
The graphs (a) and (c) are equivalent 
under local complementation without any additional permutation. 
{(d)} As shown in Ref.~\cite{Ji2010_LU-LC}, after LU transformations 
on the graph (c) a single edge (shown in red) can be created. But the graphs
(c) and (d) are not LC equivalent. Now, (b) and (d) are equivalent 
under local complementation, without any permutation.
\label{fig:nicegraphs1}
}
\end{figure}

Indeed, with our methods we can directly construct the known 
counterexample and understand why it is a counterexample. Consider
the graph with 27 vertices shown in Fig.~\ref{fig:tripartite-construction}. 
This is a slight modification of the graph in Fig.~\ref{fig:bipartite-construction}
discussed before. It is a bipartite graph with six vertices (the central vertices)
in one part of the bipartition and 21 vertices (the left and the right vertices) 
in the other part. As before, one can directly  verify the following facts: 

\begin{itemize}

\item If we apply $X^\frac{1}{4}$ on all the $21$ qubits on the right-hand 
side and one the left-hand side, any possible two-edge between the central 
qubits will be created with weight one. Higher-order edges will not be created. 
In addition, any single-qubit edge on the central qubits will occur with weight
$15/4$, but this can be corrected by local $Z$ gates on the central qubits. 

\item The number of created two-edges on the central qubits is $15=\binom{6}{2}$,
which is odd. This cannot be created by local complementation using qubits on
the right side or the left side, since each of these local complementations
changes an even number of edges (either $10=\binom{5}{2}$ or $6=\binom{4}{2}$)
on the central qubits. 

\item
Finally, the Lemma from the previous section proves that it is sufficient to
consider these special local complementations. 
\end{itemize}
So, the graph in 
Fig.~\ref{fig:tripartite-construction} is a counterexample to the LU-LC 
conjecture. 
The graph before and after this transformation is also shown in 
Fig.~\ref{fig:nicegraphs1} (a) and (b).

It remains to show that this example is equivalent to the example from 
Ref.~\cite{Ji2010_LU-LC}.
The adjacency matrix is explicitly given in that reference, so one may be tempted 
to check the LC equivalence directly by solving the equations over the binary field
as outlined in the previous section \cite{nestgraphicaldescription, martenefficient}. 
This, however, is not feasible, as there are $27! \approx10^{28}$ permutations of 
the vertices which have to be taken into account. Nevertheless
one can find the sequence of local complementations and permutations 
using the following ideas: 

\begin{itemize}

\item The first observation is that the graph from Ref.~\cite{Ji2010_LU-LC} 
is bipartite, also with six vertices on the one side and 21 vertices on the 
other side. 

\item The second observation is that certain types of sequences of local 
complementations of the graph from Fig.~\ref{fig:tripartite-construction} 
do not change the property that it is bipartite, but the partition changes. For 
instance, if we perform a local complementation on the six central vertices, 
then the local complementation on some other vertices $\{a,b,c,\dots\}$ on 
the other side, then local complementation on the six central vertices again, 
and finally on the vertices $\{a,b,c,\dots\}$ again, the graph typically stays
bipartite (again with a $6$ vs.~$21$ splitting), but the qubits are permuted
and the two parts have changed.

\item Considering such transformations, one finds with some exhaustive search
transformations which transform the graph from Fig.~\ref{fig:tripartite-construction} 
to a graph with the same degree distribution as the graph from Ref.~\cite{Ji2010_LU-LC}.
More specifically, the set $\{a,b,c,\dots\}$ consists of six qubits, four on the right
side and two on the left side. 
For these transformations and with the help of the given degree distribution one finds 
a final permutation to map one graph to the other.

\end{itemize}

With this method we were able to show that under local complementation
the initial state in Ref.~\cite{Ji2010_LU-LC} is equivalent to our state 
from Fig.~\ref{fig:tripartite-construction}, and the final state from 
Ref.~\cite{Ji2010_LU-LC} is also equivalent to our final state. The 
corresponding graphs are also displayed in Fig.~\ref{fig:nicegraphs1}
and a Mathematica file with the explicit calculation is contained in 
the supplementary information \cite{arxivweb}. This proves that
our method explains the numerically found counterexample from Ref.~\cite{Ji2010_LU-LC}. 

The example with 27 qubits is the smallest example that we could find
with our methods. As stated in Ref.~\cite{Ji2010_LU-LC}, also the numerical
search gives as a minimal example the example with 27 qubit. It would be 
interesting to decide whether this is indeed the smallest possible example. 

Finally, we should add that further counterexamples to the LU-LC conjecture 
have been found by M. Grassl and B. Zeng \cite{grasslcom}. They found 
counterexamples for $N \in \{ 27, $ $28,$ $ 35, 36, 38, 39, 40, 41, 42, 43, $ $ 44, 
45, 46, 47, 48, 49, $ $ 50, 51, 52, 54, 55, 56, 59, 60,$ $ 76, $ $115, 116, 123, 124, 
131, 132 \}$
using randomized combinatorial search and methods from coding theory. With 
our methods, we were not able to find counterexamples for all of these values of 
$N$. This suggests that our construction is not the only way
for generating counterexamples and more refined construction methods
may be found.

\section{Conclusion}
In summary, we have investigated the action of local unitary transformations 
beyond local Clifford operations on graph states and hypergraph states. For 
the action of $X_i^\alpha$ gates we found a graphical rule in terms of weighted
hypergraphs. Using this rule, we showed an example where hypergraph states and 
graph states are locally equivalent. We also provided a method to generate 
systematically counterexamples to the LU-LC conjecture. This also allowed to 
understand a previously known counterexample. 

For further research, there are several open questions. First, it would be 
interesting to explore the application of our rule to hypergraph states. For
instance, it has been suggested that if one has two hypergraph states where one
has only $k$-edges and the other has only $\ell$-edges then they may be locally 
inequivalent \cite{rossinjp}. Our graphical rule may be useful to find 
counterexamples to this question. Another interesting problem is to find 
graphical rules for other local 
transformations. They may not be applicable in all situations, as the transformations
may, in general, lead out of the space of weighted hypergraph states. Nevertheless, 
for a restricted class of states such transformations and rules may be possible.

\section{Acknowledgements}
We thank Mariami Gachechiladze, Markus Grassl, and David Gross for discussions. 
This work has been supported by the DFG and the ERC (Consolidator 
Grant 683107/TempoQ).



\section*{References}
\begin{thebibliography}{99}

\bibitem{heinpra}
M.~Hein, J.~Eisert and H.~J.~Briegel, Phys. Rev. A {\bf 69}, 062311 (2004).

\bibitem{hein}
M. Hein, W. D\"ur, J. Eisert, R. Raussendorf, M. Van den Nest, and H.-J. Briegel,
{\it Entanglement in Graph States and its Applications},
in {\em Quantum Computers, Algorithms and Chaos}, edited by G.
Casati, D.L. Shepelyansky, P. Zoller, and G. Benenti (IOS Press,
Amsterdam, 2006), quant-ph/0602096.

\bibitem{rossinjp}
M.~Rossi, M.~Huber, D.~Bru\ss, and C.~Macchiavello,
New J. Phys. {\bf 15}, 113022 (2013).

\bibitem{Qu2013_encoding}
R.~Qu, J.~Wang, Z.~Li, and Y.~Bao,
Phys. Rev. A {\bf 87}, 022311 (2013).

\bibitem{scripta}
M. Rossi, D. Bru{\ss}, and C. Macchiavello,
Phys. Scr. {\bf T160}, 014036 (2014). 

\bibitem{guehnejpa}
O. G\"uhne, M. Cuquet, F. E. S. Steinhoff, T. Moroder, M. Rossi,
D. Bru{\ss}, B. Kraus, and  C. Macchiavello,
J. Phys. A: Math. Theor. {\bf 47}, 335303 (2014).

\bibitem{chenlei}
X.-Y. Chen and L. Wang, 
J. Phys. A: Math. Theor. {\bf 47}, 415304 (2014).

\bibitem{lyons}
D. W. Lyons, D. J. Upchurch, S. N. Walck, and C. D. Yetter,
J. Phys. A: Math. Theor. {\bf 48}, 095301 (2015).

\bibitem{lyonsnew} 
D. W. Lyons, N. P. Gibbons,
M. A. Peters, D. J. Upchurch, S. N. Walck, and E. W. Wertz, 
arXiv:1609.01306.

\bibitem{gbg}
M. Gachechiladze, C. Budroni, and O. G\"uhne,
Phys. Rev. Lett. {\bf 116}, 070401 (2016).

\bibitem{graphapp2} D. Schlingemann and R.F. Werner, 
Phys. Rev. A {\bf 65}, 012308 (2002).

\bibitem{grassl}
M. Grassl, A. Klappenecker, and M. R\"otteler,
Proceedings 2002 IEEE International Symposium on Information Theory (ISIT 2002),
Lausanne, Switzerland, June/July 2002, p. 45, 
quant-ph/0703112. 

\bibitem{mqc}
H.~J. Briegel, D.~E. Browne, 
W. D\"ur, R. Raussendorf, and 
M. Van den Nest,
Nat. Phys. {\bf 5}, 19 (2009).

\bibitem{adan8pla}
A. Cabello, A.J. Lopez-Tarrida, P. Moreno, and J.R. Portillo,
Phys. Lett. A {\bf 373}, 2219 (2009);
Erratum: 
Phys. Lett. A {\bf 374}, 3991 (2010).

\bibitem{adan8pra}
A. Cabello, A.J. Lopez-Tarrida, P. Moreno, and J.R. Portillo,
Phys. Rev. A {\bf 80}, 012102 (2009).

\bibitem{nestgraphicaldescription}
M.~van~den~Nest, J.~Dehaene and B.~De~Moor, 
Phys. Rev. A {\bf 69}, 022316 (2004).

\bibitem{martenefficient}
M.~van~den~Nest, J.~Dehaene and B.~De~Moor, 
Phys. Rev. A {\bf 70}, 034302 (2004).

\bibitem{maartenlulc}
M.~van~den~Nest, J.~Dehaene and B.~De~Moor, 
Phys. Rev. A {\bf 71}, 062323 (2005).

\bibitem{zengpra}
B. Zeng, H. Chung, A. W. Cross, and I. L. Chuang,
Phys. Rev. A {\bf 75}, 032325 (2007).

\bibitem{nestgross}
D. Gross and M. Van den Nest,
Quantum Inf. Comp. {\bf 8}, 263 (2008).

\bibitem{Ji2010_LU-LC}
Z. Ji, J. Chen, Z. Wei, and M. Ying,
Quantum Inf. Comp. {\bf 10}, 97 (2010).

\bibitem{carolinapra}
C. Kruszynska and B. Kraus,
Phys. Rev. A {\bf 79}, 052304 (2009).

\bibitem{Carle2013}
T.~Carle, B.~Kraus, W.~D\"ur, and J.I.~de Vicente,
Phys. Rev. A {\bf 87}, 012328 (2013).

\bibitem{gtg}
M. Gachechiladze, N. Tsimakuridze, and O. G\"uhne,
J. Phys. A: Math. Theor. {\bf 50}, 19LT01 (2017). 

\bibitem{arxivweb} The file is also available in the submission files of 
this paper on the arxiv, see {\tt https://arxiv.org/e-print/1611.06938v1.}

\bibitem{grasslcom}
M. Grassl, private communication, 2016.

\end{thebibliography}
\end{document}